\documentclass[12pt,a4paper]{article}%
\usepackage{amsmath}
\usepackage{amsfonts}
\usepackage{amssymb}
\usepackage{graphicx}%
\newtheorem{theorem}{Theorem}

\newtheorem{notation}[theorem]{Notation}

\newtheorem{remark}[theorem]{Remark}

\begin{document}

\title{Synthesizing the L\"{u} attractor by parameter-switching}
\author{Marius-F. Danca\\$^{a}$Department of Mathematics and Computer Science, \\Avram Iancu University, \\400380 Cluj-Napoca, Romania; \\$^{b}$Institute of Science and Technology,\\400487 Cluj-Napoca, Romania}
\date{}
\maketitle

\begin{abstract}
In this letter we synthesize numerically the L\"{u} attractor starting from
the generalized Lorenz and Chen systems, by switching the control parameter
inside a chosen finite set of values on every successive adjacent finite time
intervals. A numerical method with fixed step size for ODEs is used to
integrate the underlying initial value problem. As numerically and
computationally proved in this work, the utilized attractors synthesis
algorithm introduced by the present author before, allows to synthesize the
L\"{u} attractor starting from any finite set of parameter values.

\end{abstract}


\section{Introduction}

Consider the following unified chaotic system (bridge between the
Lorenz and Chen systems) \cite{Lu}:
\begin{equation}
\begin{array}
[c]{l}
\overset{.}{x}_{1}=(25p+10)(x_{2}-x_{1}),\\
\overset{.}{x}_{2}=(28-35p)x_{1}+(29p-1)x_{2}-x_{1}x_{3},\\
\overset{.}{x}_{3}=x_{1}x_{2}-\left(  8+p\right)  /3x_{3},
\end{array}
\label{unified}
\end{equation}

\smallskip\noindent\noindent where the parameter $p\in\left[  0,1\right]  .$
As it is known now, for $p\in\lbrack0,0.8)$~(\ref{unified}) models the
canonical Lorenz system \cite{Celikovski and Chen}, for $p=0.8$ the system
becomes L\"{u} system \cite{Lu and Chen}, while when $p\in(0.8,1]$, the system
becomes Chen system \cite{Chen and Ueta}. Therefore, this system is likely to
be the simplest chaotic system bridging the gap between the Lorenz and the
Chen systems.

The above three systems share some common properties such as: they all have
the same symmetry, dissipativity, stability of equilibria, similar
bifurcations and topological structures and belong to the generalized Lorenz
canonical family \cite{Celikovski and Chen}.

In the mentioned references, a positive answer to the question as if it is
possible to realize a continuous transition from one to another system is given.

In this letter, we present a discontinuous transition algorithm between the
Lorenz and the Chen systems with whitch the L\"{u} attractor can be
synthesized. For this purpose, the parameter switching method introduced in
\cite{Danca et al 1} is utilized.

The present work is organized as follows: Section 2 presents the synthesis
algorithm, while in Section 3 the L\"{u} attractor is synthesized in both
deterministic and random ways via the mentioned synthesis algorithm.

\section{Attractors synthesis algorithm}

Consider a class of dissipative autonomous dynamical systems modeled
by the following initial value problem:
\begin{equation}
S:~\dot{x}=f_{p}(x),\quad x(0)=x_{0}, \label{ivp general}
\end{equation}
\noindent where $p\in\mathbb{R}$ and
$f_{p}:\mathbb{R}^{n}\longrightarrow \mathbb{R}^{n}\,$ has the
expression
\begin{equation}
f_{p}(x\mathbf{)=}g(x\mathbf{)}+pMx, \label{2}
\end{equation}

\noindent with $g:\mathbb{R}^{n}\longrightarrow\mathbb{R}^{n}~\ $being a
vector continuous nonlinear function, $M~$a real constant $n\times n$
matrix,$~x_{0}\in\mathbb{R}^{n}$, and the maximal existence
interval$~I=[0,\infty).$

For the L\"{u} system (\ref{unified}), one has:
\[
M=\left(
\begin{array}
[c]{ccc}%
0 & 25 & 0\\
-35 & 29 & 0\\
0 & 0 & -1/3
\end{array}
\right)  ,~g(x)=\left(  10\left(  x_{2}-x_{1}\right)  ,~x_{2}-x_{1}
x_{3}+28x_{1},x_{1}x_{2}-8/3x_{3}\right)  ^{T}
\]

\noindent with divergence $div~f_{p}(x)<0$ for$~p\in\lbrack0,1],$ so the
system is dissipative.

The existence and uniqueness of solutions on the maximal existence interval
$I~$are assumed. Also, without any restriction, it is supposed that
corresponding to different $p$, there are different global attractors. Because
of numerical characteristics of the attractors synthesis (AS) algorithm and
for sake of simplicity, by a \emph{(global) attractor} in this letter one
understand without a significant loss of generality, only the approximation of
the $\omega$-\emph{limit set}, as in \cite{Foias}, is plotted after neglecting
a sufficiently long period of transients (for background about attractors see
\cite{milnor}).

\begin{notation}
Let $\mathcal{P\subset}\mathbb{R}$ be the set of all admissible values for $p$
and $\mathcal{A}$ the set of all corresponding global attractors, which
includes attractive stable fixed points, limit cycles and chaotic attractors.
Also, denote by $\mathcal{P}_{N}$ a finite subset of $\mathcal{P}$ for some
positive integer $N>1$ and the corresponding subset of attractors
$\mathcal{A}_{N}\subset\mathcal{A}.$
\end{notation}

Because of the assumed dissipativity, $\mathcal{A}$ is a non-empty set.
Therefore, following the above assumptions, a bijection between $\mathcal{P}$
and $\mathcal{A},$ $F:$ $\mathcal{P\rightarrow A},$ can be considered. Thus,
to each $p\in\mathcal{P}$ corresponds a unique global attractor $A_{p}%
\in\mathcal{A}$ and conversely for each global attractor there exists a unique
parameter value $p\in\mathcal{P}$.

In \cite{Danca et al 1}, it is proved numerically that switching,
indefinitely in some periodic way,the parameter $p$ inside
$\mathcal{P}_{N}$ over finite time subintervals, while (\ref{ivp
general}) is integrated with some numerical method for ODEs with
fixed step size $h$, any attractor of $\mathcal{A}_{N}$ can be
synthesized. For a chosen $N$, consider $I$ being partitioned in to
consecutive sets of $N~$finite adjacent time subintervals $I_{i}$, :
$I=(I_{1}\cup I_{2}\cup\ldots\cup I_{N})\cup(I_{1}\cup
I_{2}\cup\ldots\cup I_{N})\cup\ldots$ of lengths $\Delta
t_{i},~i=1,2\ldots,N$. If, in each subinterval $I_{i},$ while some
numerical method with single fixed step size $h$ integrates
(\ref{ivp general}), $p$ is switched as follows: $p=p_{i},~$for
$t\in I_{i},$ Then, \noindent a \emph{synthesized attractor},
denoted by $A^{\ast},$ can be generated$.$ The simplest way to
implement numerically the AS algorithm is to choose $\Delta t_{i}$
as a multiple of $h$. Thus, the AS algorithm can be symbolically
written for a fixed step size $h$ as follows:
\begin{equation}
\lbrack m_{1}p_{1},~m_{2}p_{2},\ldots,m_{N}p_{N}], \label{SA}
\end{equation}

\noindent where $m_{k}$ are some positive integers (weights) and by
$m_{k}p_{k}$ one understands that in the $k$-th time subinterval $I_{k},~$of
length $m_{k}h,$ $p~$receives the value $p_{k}.$

In \cite{Danca et al 1}, it is proved numerically that $A^{\ast}$ is
\emph{identical}\footnote{Identicity is understood in a geometrical
sense: two attractors are considered to be (almost) identical if
their trajectories in the phase space coincide. The word
\emph{almost} corresponds to the case of chaotic attractors, where
identity may appear only after infinite time. Supplementarily,
Poincar\'{e} sections and Haussdorf distance between trajectories
are utilized to underline this identity.} to $A_{p^{\ast}}$ for

\begin{equation}
p^{\ast}=\frac{\sum\limits_{k=1}^{N}m_{k}p_{k}}{\sum\limits_{k=1}^{N}m_{k}
}\text{.} \label{p formula}
\end{equation}

\noindent

For example, the sequence $\left[  1p_{1},2p_{2}\right]  $ means that
$m_{1}=1,$ $m_{2}=2$ and the synthesized attractor $A^{\ast}~$is synthesized
as follows: in the first time interval $I_{1}~$of length $\Delta t_{1}=h,$ the
numerical method solves (\ref{ivp general}) with $p=p_{1};$ next, for the
second time interval $I_{2}$ of length $\Delta t_{2}=2h$, $p=p_{2},$ and the
algorithm repeats. If we apply this scheme to (\ref{unified}) for $p_{1}=0.8$
(chaotic L\"{u} attractor) and $p_{2}=0.959$ (chaotic Chen attractor), one
obtains the synthesized regular Chen attractor $A^{\ast}$ which is identical
to $A_{p^{\ast}}~$with $p^{\ast}=\left(  p_{1}+2p_{2}\right)  /3=0.906~$in
(\ref{p formula}), corresponding to a stable periodic limit cycle. In Fig. 1,
to underline the identity, phase plots, time series, histograms and
Poincar\'{e} sections superimposed were utilized beside Haussdorf distance
between the two attractors which is of order $10^{-2}\div10^{-3}~$conferring a
good accuracy to AS.

It is noted that the AS algorithm can be applied even in some random way:
because $p^{\ast}~$in (\ref{p formula}) is defined in a convex manner (if
denoting $\alpha_{k}=m_{k}/\sum\limits_{k=1}^{N}m_{k}<1,$ then$~p^{\ast}%
=\sum\limits_{k=1}^{N}\alpha_{k}p_{k},$ with$\ \sum\limits_{k=1}^{N}\alpha
_{k}=1)$ and based on the bijective function $F$, any synthesized attractor
$A^{\ast}$ is located inside the set $\mathcal{A}_{N}$ (all elements, i.e.
attractors, are ordered with the order endowed by $F)~$and whatever (random)
scheme (\ref{SA}) is used, the result is the same \cite{Danca1}.

\noindent The random AS can be implemented e.g. by generating a
sequence (\ref{SA}) with a random uniform distribution of $p$
\cite{Danca1} which is supposed to generate all the integers
$1,\ldots,N~$(Fig. 2)$.~$

\noindent Now, $p^{\ast}$ is given by the following formula:
\begin{equation}
p^{\ast}=\frac{\sum\limits_{i=1}^{N}m_{i}^{^{\prime}}p_{i}}{\sum
\limits_{i=1}^{N}m_{i}^{^{\prime}}} \label{p random}
\end{equation}

\noindent where $m_{i}^{\prime}$ counts the number of $p_{i\text{ }}$.
Obviously, now, $I$ has to be chosen large enough, such that (\ref{p random})
can converge to $p^{\ast}$ (the precise value in this case for $p^{\ast}$
could be obtained only for $I=[0,\infty)$ ).

\begin{remark}
i) The AS algorithm is useful in the applications where some $p$ are not
directly accessible.\newline ii) The AS algorithm can be viewed as an
explanation for the way regular or chaotic behaviors may appear in natural
systems. \newline ii) Being a numerical algorithm, AS has limitations. For
example, for relatively large switches of $p$ or $m,$ or for a too-large
number $N,$ $A^{\ast}$ could present some "corners". Also, obviously, the $h$
size may influence the AS algorithm performances (ideally, $\ h$ should
decrease to zero). Some details and other related aspects about the errors can
be found in \cite{Danca et al 1} and \cite{Danca1}). \newline iii) In the
general case of a dynamical system modeled by (\ref{ivp general}), the only
restriction to synthesize a chaotic attractor, when starting from regular
attractors, is that inside the set $\mathcal{A}_{N}$ there are chaotic
attractors (and vice-versa for regular synthesized attractors).\newline iv)
The AS can be used as a kind of control-like method \cite{Danca 2} or
anticontrol \cite{Danca et al 1}.\newline v) Near several continuous dynamical
systems (such as the Chen system, R\"{o}ssler system, Rabinovich-Fabrikant
system (\cite{xiaodong et al}) minimal networks, Lotka-Volterra system, L\"{u}
system, Rikitake system), the AS algorithm was also applied successfully to
systems of fractional orders \cite{Danca Kai}.
\end{remark}

\section{\smallskip L\"{u} attractor synthesis}

The numerical results in this section are obtained using the standard
Runge-Kutta algorithm with fixed integration time step $h=0.001$.

To visualize how the AS works, the bifurcation diagram was plotted (Fig. 3).
Next, we synthesize the L\"{u} attractor starting from different values for
$p$ and using deterministic or random schemes (\ref{p formula}). In this
simulation, once we fixed $N,$ all we need is to choose $m$ and $\mathcal{P}%
_{N}~$so that the equation (\ref{p formula}) with $p^{\ast}=0.8$
corresponding to the L\"{u} attractor, can be verified. Besides
Poincar\'{e} sections and histograms, Haussdorf distance between
$A^{\ast}~$and $A_{p^{\ast}}$ (\cite{Falconer} p.114) was computed,
in this case, in order of $10^{-2} \div10^{-3},$ which indicates a
good approximation.

First we applied the deterministic scheme (\ref{SA}) for $N=2,$ $p_{1}=0.2$
(corresponding to generalized the Lorenz system, Fig. 4 a), $\ $and$~p_{2}=1$
(corresponding to the Chen system, Fig. 4 b) with the scheme $[1p_{1},3p_{2}%
]$. In this case, the synthesized attractor $A^{\ast}$ is identical to
$A_{p^{\ast}}$ with $p^{\ast}=0.8=\left(  1p_{1}+3p_{2}\right)  /4$ (Fig. 4
c). In Fig. 4 d and e, the histograms and Poincar\'{e} sections of both
attractors, $A^{\ast}$ and $A_{p^{\ast}},~$are plotted superimposed to
underline the identity.

Because the solution of (\ref{p formula}) for given $N,~P_{N}$ is not unique,
the L\"{u} attractor can be obtained in, theoretically, infinitely many ways.
Thus, we chose $N=5$, $p_{1}=0.47,~p_{2}=0.585,~p_{3}=0,678~$(corresponding to
the Lorenz system)$,~p_{4}=0.905,$ $p_{5}=0.9405$ (corresponding to the Chen
system)~and $\ m_{1}=m_{2}=m_{3}=m_{4}=1,$ $m_{5}=4,$ again $p^{\ast
}=0.8=(p_{1}+p_{2}+p_{3}+p_{4}+4p_{5})/8.$ $A_{p_{1,...,5}}$ and $A^{\ast},$
$A_{p^{\ast}}$ are presented in Fig. 5 with Poincar\'{e} sections and histograms.

Using the random way presented in Fig. 2, the L\"{u} attractor can be
synthesized with, for example, $p_{1}=0.6$ and $p_{2}=1$ (Fig. 6).

\section{Conclusion}

The design AS algorithm has been utilized to generate numerically the L\"{u}
attractor starting from his "neighbors", the Lorenz and Chen attractors, not
by continuous transformations as before but by discontinuous parameter
switching inside a chosen parameter set.

\newpage

\begin{figure}
\begin{center}
\includegraphics[clip,width=1\textwidth]{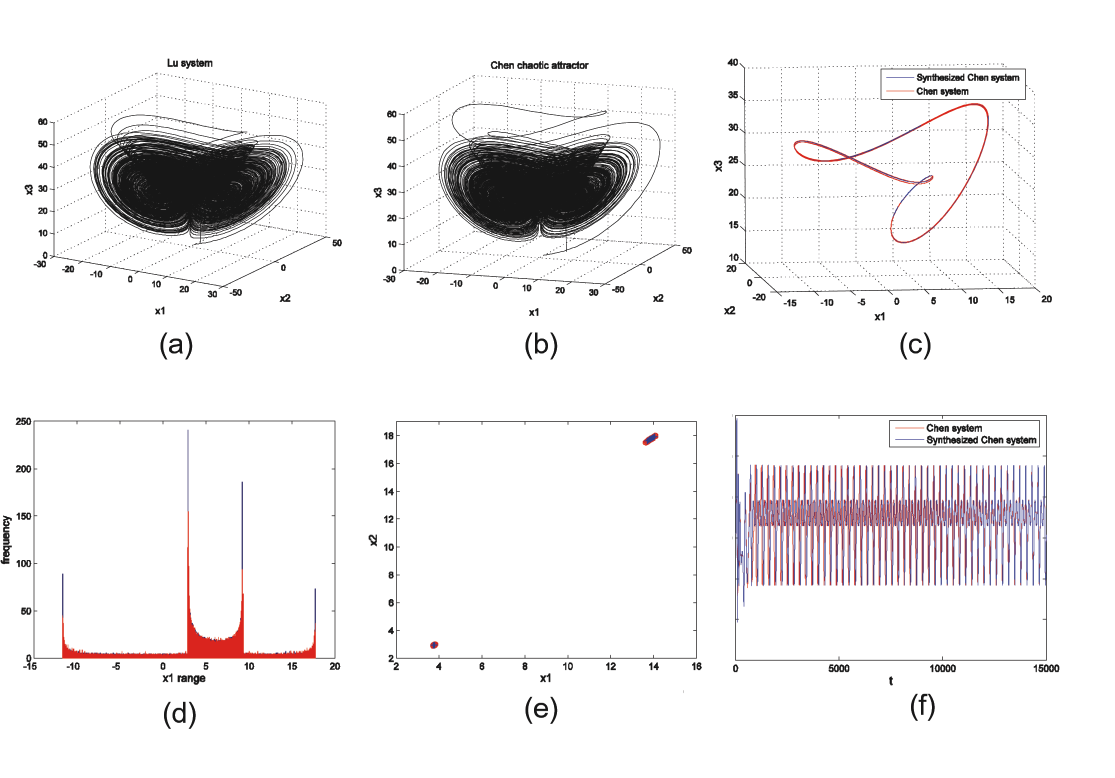}
\caption{Synthesis of a stable limit cycle for the Chen
attractor$,~$obtained using the scheme $[1p_{1},2p_{2}]$ with
$p_{1}=0.8$ and $p_{2}=0.959$ and $p^{\ast}=0.906$; a) L\"{u}
attractor; b) Chen attractor; c) $A^{\ast}$ and $A_{p^{\ast}}$
plotted superimposed; d) Histograms of $A^{\ast }$ and
$A_{p^{\ast}}~$superimposed; e) Poincar\'{e} superimposed sections
with plane $x_{3}=28$ of $A^{\ast}$ and $A_{p^{\ast}}.~$f) Time
series with transients of component $x_{1}$ of $A^{\ast}$ and
$A_{p^{\ast}}$ superimposed.}
\end{center}
\end{figure}

\begin{figure}
\begin{center}
\begin{equation}
\begin{array}
[c]{l}
repeat\\
~~~~label=\operatorname{rand}(N)\\
~~~~if~label=1~then\\
~~\ ~~~~~~~integrate~(\ref{ivp general})~with~p=p_{1}\\
~~~~~~~~~~inc(m_{1}^{^{\prime}})\\
~~~~if~label=2~then\\
~~~~~~~~~~integrate~(\ref{ivp general})~with~p=p_{2}\\
~~~~\ ~~~~~inc(m_{2}^{^{\prime}})\\
~~~~~\ldots\\
~~~~if~label=N~then\\
~~~~~~~~~~integrate~(\ref{ivp general})~with~p=p_{N}\\
~\ ~~\ ~~~~~inc(m_{N}^{^{\prime}})\\
~~~~~t=t+h\\
until~t\geq T_{\max}
\end{array}
\label{random}
\end{equation}
\caption{Random SA algorithm.}
\end{center}
\end{figure}

\begin{figure}
\begin{center}
\includegraphics[clip,width=1.1\textwidth]{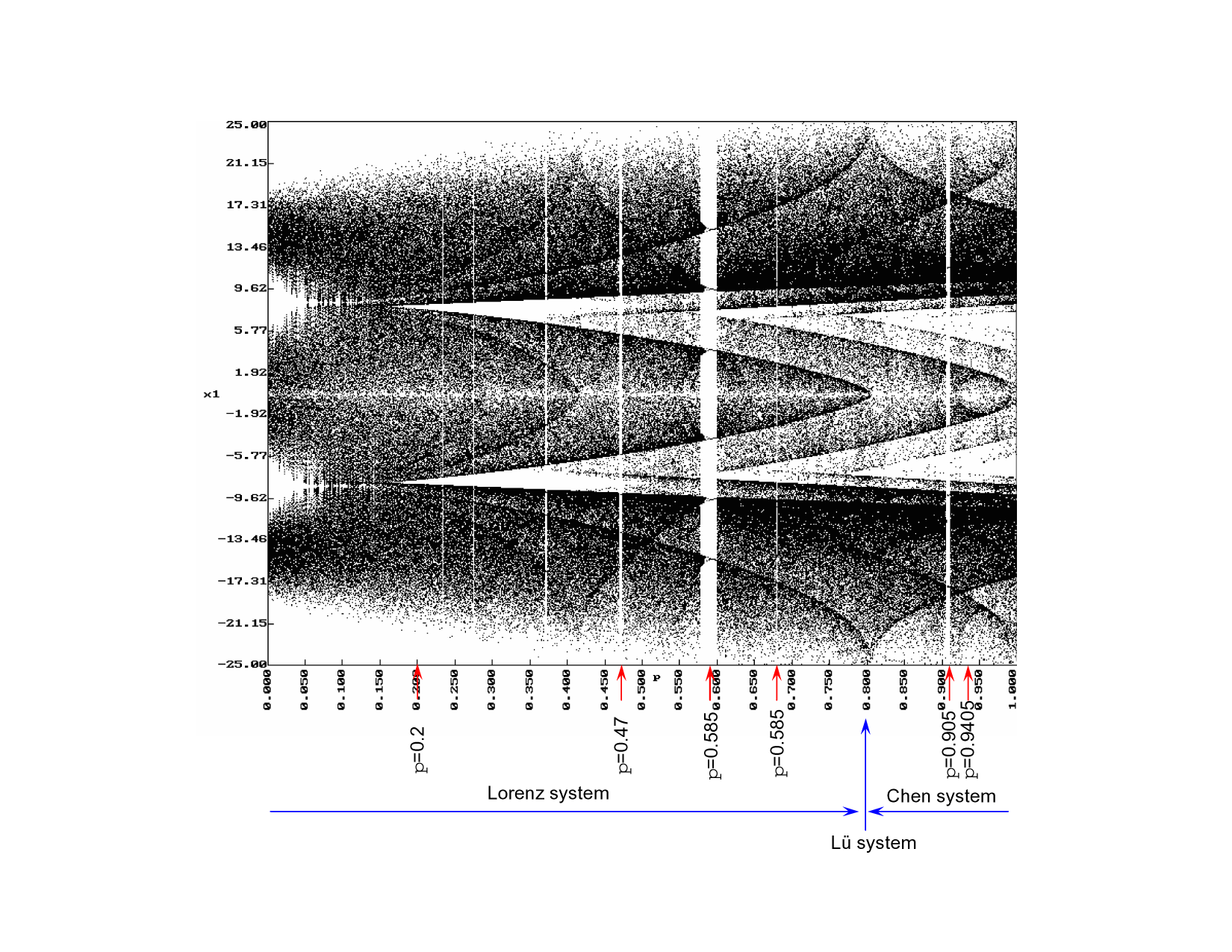}
\caption{Bifurcation diagram for the L\"{u} system.}
\end{center}
\end{figure}

\begin{figure}
\begin{center}
\includegraphics[clip,width=1\textwidth]{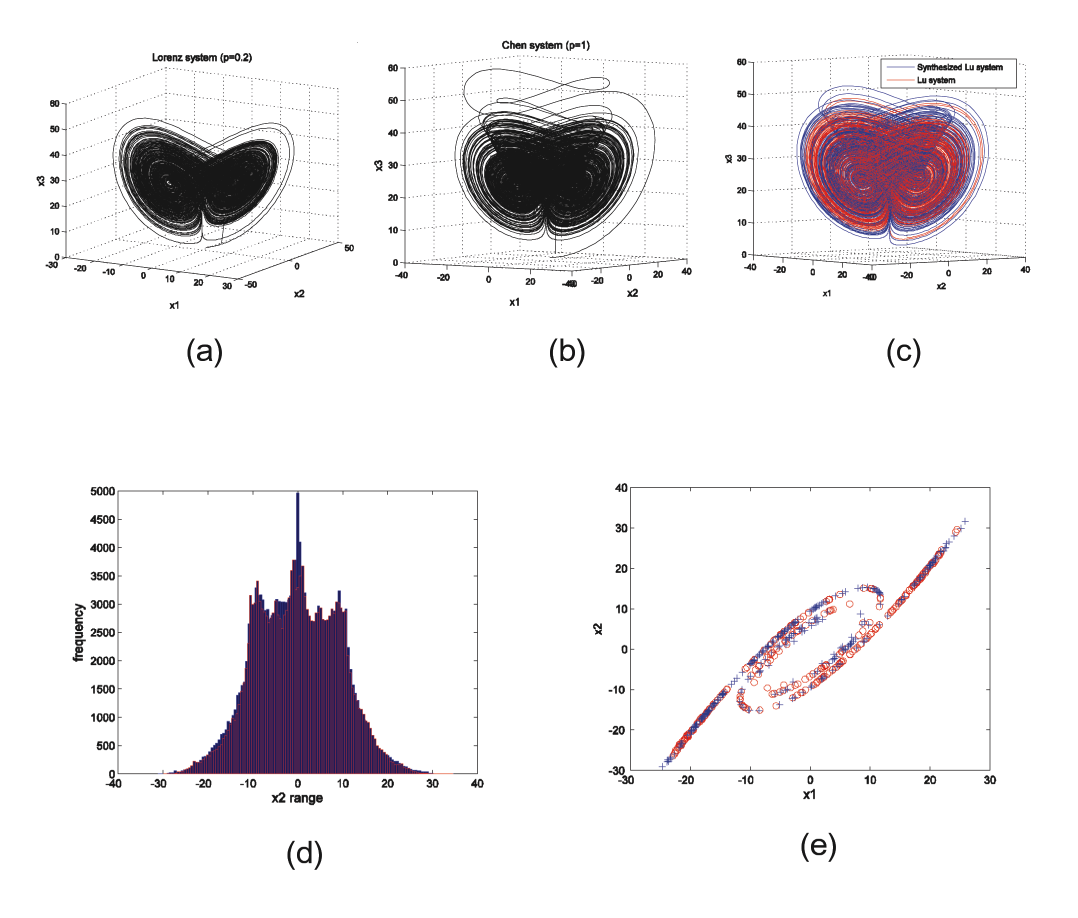}
\caption{The synthesized L\"{u} attractor obtained with scheme
$[1p_{1},3p_{2}]$ for $p_{1}=0.2,$ $p_{2}=1$; a) $A_{p_{1}}$; b)
$A_{p_{2}};~$c) $A^{\ast}$ and $A_{p^{\ast}}$ plotted superimposed;
d) Superimposed histograms; e) Superimposed Poincar\'{e} sections.}
\end{center}
\end{figure}

\begin{figure}
\begin{center}
\includegraphics[clip,width=0.95\textwidth]{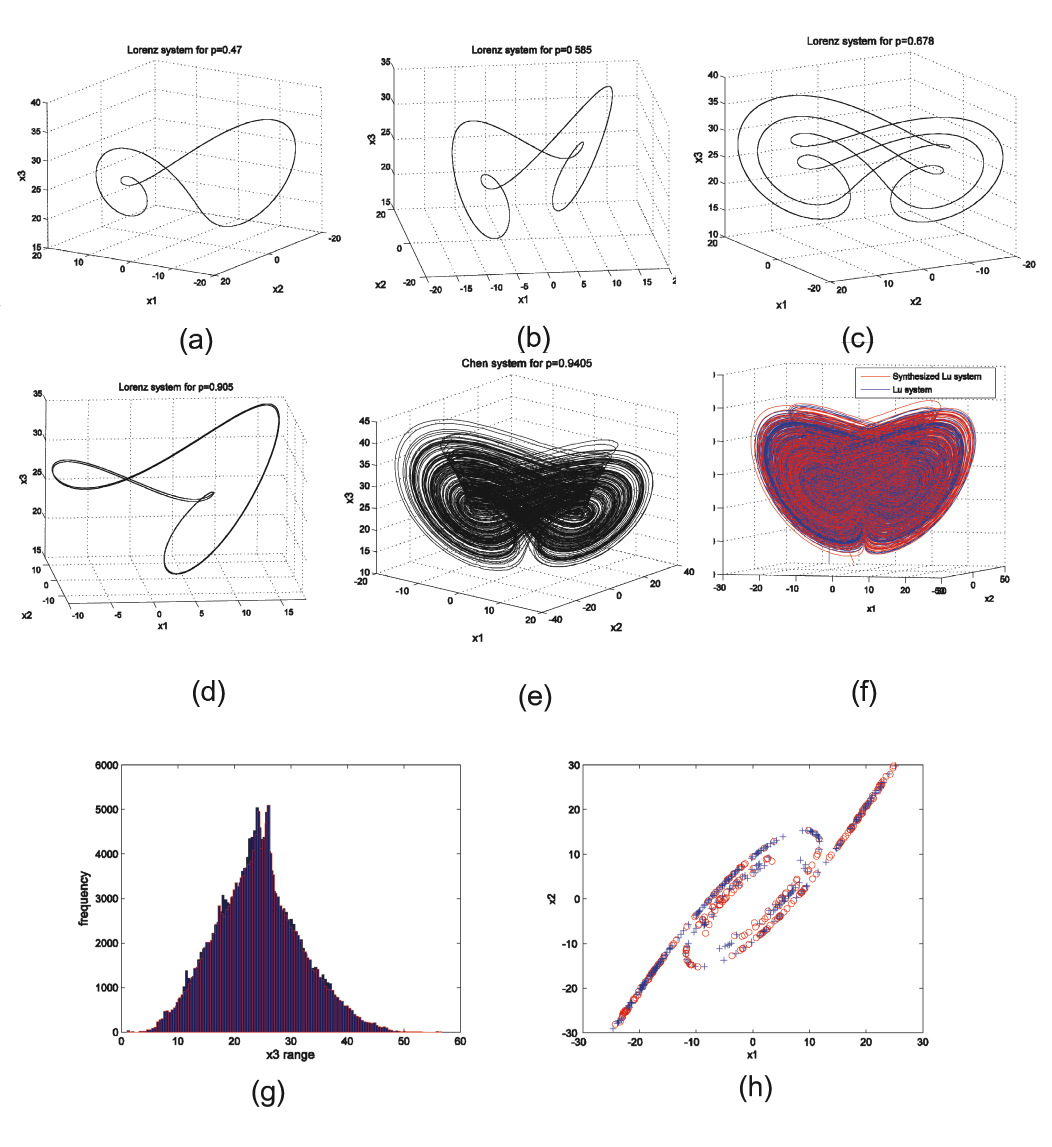}
\caption{The synthesized L\"{u} attractor obtained with scheme
$[1p_{1},$ $1p_{2},1p_{3},1p_{4},4p_{5}]$ for
$p_{1}=0.47,~p_{2}=0.585,~p_{3}=0,678,$ $p_{4}=0.905$ and
$p_{5}=0.9405.$ a-e) Attractors $A_{p_{i}},$ $i=1,\ldots,5;$ f)
$A^{\ast}$ and $A_{p^{\ast}}~$plotted superimposed; g) Superimpose
histograms; h) Superimposed Poincar\'{e} sections. }
\end{center}
\end{figure}

\begin{figure}
\begin{center}
\includegraphics[clip,width=1\textwidth]{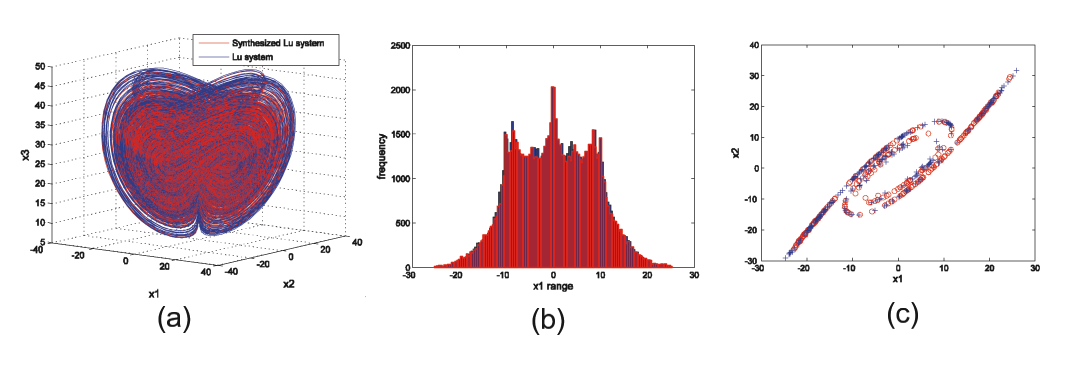}
\caption{The synthesized L\"{u} attractor obtained with the random
scheme in Fig. 1 with uniform distribution of values $p_{1}=0.6$
and$~p_{2}=1.$a) Superimposed histograms; b) Superimposed
Poincar\'{e} sections. }
\end{center}
\end{figure}

\end{document}